\documentclass[%
aip,
amsmath,amssymb,
reprint,%
]{revtex4-1}

\usepackage{amssymb,amsmath,graphicx,amsthm,wrapfig,epstopdf,esint,afterpage,marginnote,afterpage,lipsum,color,tabularx}
\usepackage{dcolumn,bm}
\usepackage[dvipsnames]{xcolor}
\usepackage{natbib}
\usepackage[version=4]{mhchem}
\usepackage{color,sidecap,fancyhdr}
\usepackage[colorlinks]{hyperref}
\usepackage{subfigure}
\usepackage{float}
\usepackage[dvipsnames]{xcolor}
\usepackage[normalem]{ulem}

\begin{document}

\title{Time-dependent relations between gaps and returns in a Bitcoin order book}

\author{Roberto Mota Navarro}
 \email{robertomota@fisca.unam.mx}
\affiliation{Instituto de Ciencias F\'isicas, Universidad Nacional Aut\'onoma de M\'exico, Cuernavaca, Morelos, 62210, México.}
\author{Paulino Monroy Castillero}
\email{monroyca@post.bgu.ac.il}
\affiliation{
 Department of Solar Energy and Environmental Physics, Swiss Institute for Dryland Environmental and Energy Research, Blaustein Institutes for Desert Research (BIDR), Ben-Gurion University of the Negev, Sede Boqer Campus, 8499000 Midreshet Ben-Gurion, Israel
}
\author{Francois Leyvraz}
\email{leyvraz@fisca.unam.mx}
\affiliation{Instituto de Ciencias F\'isicas, Universidad Nacional Aut\'onoma de M\'exico, Cuernavaca, Morelos, 62210, México.}
\date{\today}

\begin{abstract}
Several studies have shown that large changes in the returns of an asset are associated with the sized of the gaps present in the order book In general, these associations have been studied without explicitly considering the dynamics of either gaps or returns. Here we present a study of these relationships. Our results suggest that the causal relationship between gaps and returns is limited to instantaneous causation. 
\end{abstract}
\maketitle

\section{Introduction}\label{section-introduction}
The study of financial markets has revealed that some of the properties of the price formation process can be explained to a great extent as a consequence of the market structures upon which trading is conducted rather than as a consequence of the trading strategies employed by the market participants\cite{biais1995,farmer2005,goldstein2000,gould2013}. One such example are continuous double auction markets, in which trades are mediated via a limit order book.

Several studies have shown that the statistical properties of the orders stored in a limit order book have an effect on price formation. For example, an agent based model proposed by Farmer et al. \cite{farmer2005} is able to predict 96\% of the variance in the spread of 11 stocks, even though the model has only one free parameter and assumes no rationality-based strategies in the agents, showing that the structure of an order book alone plays a big role in the price formation process. Goldstein et al\cite{goldstein2000}, report that as a consequence of decreasing the minimum price variation (tick size) between orders in the NYSE, going from 1/8 to 1/16 of a dollar, the spreads and volumes reduced their sizes, favoring thus participants which consume liquidity from the book.

Other studies have shown that there are statistical relationships between the density of price levels occupied by limit orders in the book and the price fluctuations. In the agent based models presented by Tilo et al, and Cristelli et al\cite{schmitt2012,cristelli2010}, it was shown that when large gaps between occupied large levels are present in the book, the tails of the returns distribution become fat, so that large changes in price are more common. Of particular relevance to the investigation of the relationships between order book structure and price dynamics is the empirical study conducted by Farmer et al\cite{farmer2004}, which showed that contrary to the intuitive assumption of large price fluctuations being a consequence of orders with large volumes arriving to the book\cite{gabaix2003} and consuming several occupied price levels in a single event, the largest fluctuations are actually caused by the presence of large gaps in the order book.

These studies suggest that there is a relationship resembling causality between the gaps in the order book and the returns in the sense that large price fluctuations seem to appear only when large gaps are present in the book.
To reach this conclusion, the aforementioned works rely on comparisons among members of a population of order books and in the observation of the properties of the returns and gaps collected over time, without tracking the changes in gap sizes over time and the subsequent changes in returns they would induce.

Thus it is natural to wonder whether,  in addition to the results obtained at a population level already reported in the literature, one can observe the same causal relationship in a single instance of an order book as the evolution of the gaps is tracked over time. To help us answer this question, we make use of direct causal discovery methods that are designed to take into account the chronological order of the states of the gaps and returns so that we are able to determine if a cause and effect correspondence exists in this scenario. The confirmation and quantification of a causal relationships between the present state of the gaps and the future states of the returns could, in principle, allow us to build improved volatility forecasting models that take into account the granularity of the order book as a regressor variable.


In the following we will focus our analysis on a data set from the Bitcoin/USD market because extensive order book states were available freely.

The structure of the remaining sections is as follows. In section II we discuss the methods employed to represent the states of the order book as a set of multivariate time series which is then divided into non-adjacent time windows in order to extract different statistics from each of them, as well as the procedure to obtain the correlation function average. Then we describe the two causal discovery methods applied: Granger causality and Additive Noise Models (ANM).
Section III contains the results of the correlation and causality tests between the different series extracted from the data, and finally we discuss the relevance of these results and conclude on section IV.

\section{Methods}
\subsection{Data pre-processing}
We analyzed a set of order book states from the Bitcoin trading platform BTC-e containing the first 20 price levels of each side of the book at a resolution of 10 seconds per state. The data ranges from January 2015 to August 2016 but only 14 months from this time span were included in our study since we discarded months with missing data. The months discarded are June, July and September of 2015 and January and March of 2016. 

\begin{figure}
\centering
\includegraphics[width=0.3\textwidth]{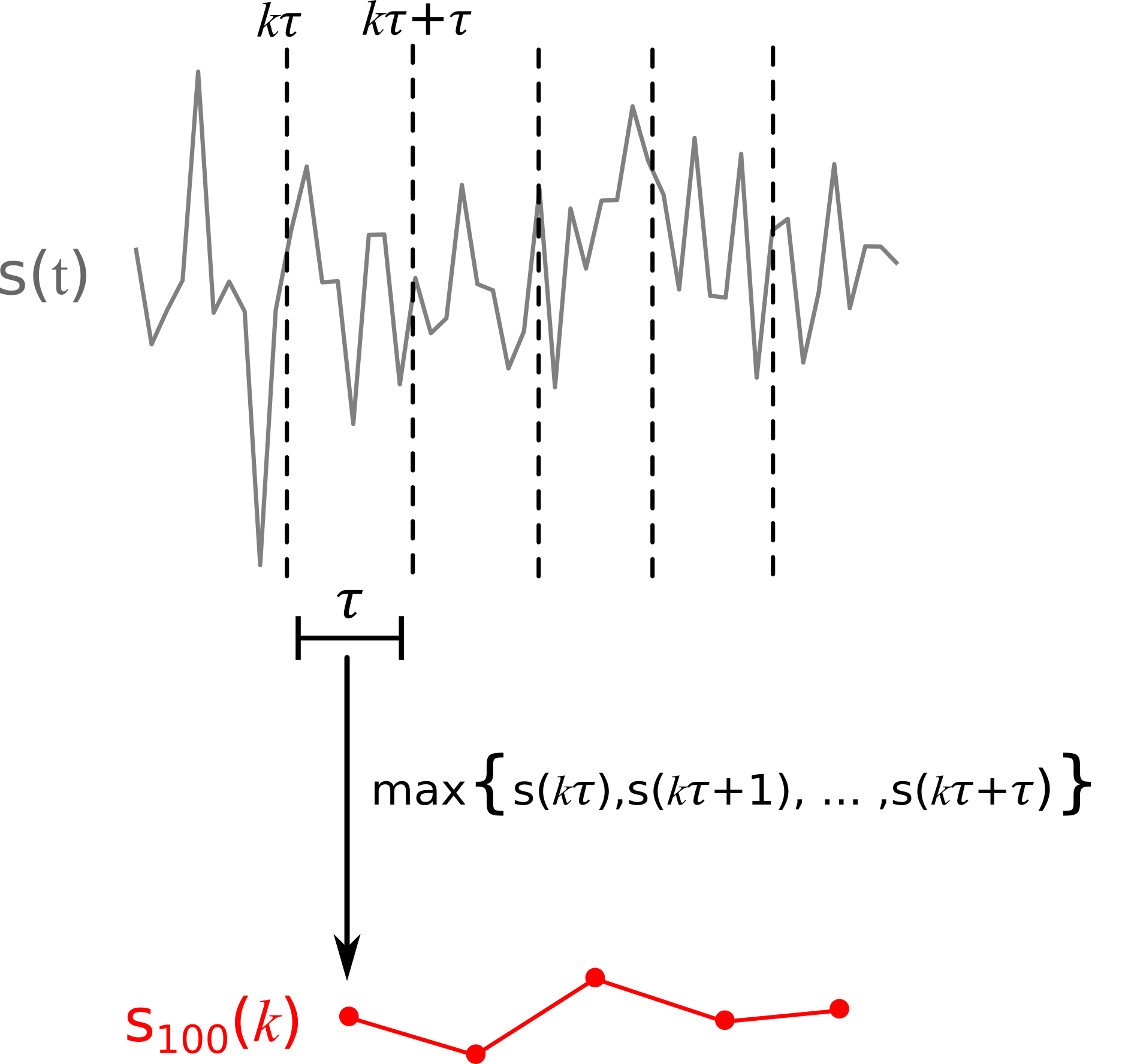}
\caption{Procedure to obtain the series of maximum (median) gaps (returns). We divide the series $s(t)$into a disjoint windows and extract from within each of these windows the maximum (median) value of $s(t)$. In the case of the gaps, the series are multivariate, containing 20 price levels per side of the book.}
\label{fig:DiagSeriesWindowReduction}
\end{figure}

The raw data of the order book can be described as follows: let $a_t^n$ ($b_t^n$) denote the $n$th ask (bid) price 
discrete time indexed by $t$, with $n=1$ corresponding to the best ask (bid). $n$ runs from $1$ to $20$
in our data set. We then define the gaps
\begin{subequations}
\begin{eqnarray}
g_1^l(t)&=\ln(&a_t^{l+1}/a_t^l)\\
g_2^l(t)&=\ln(&b_t^{l+1}/b_t^l)
\end{eqnarray}
\end{subequations}
for $l\in \{1,2,...,19\}$. We then choose a window size $\tau$ and define the time series of the maximum gaps (that is, the 100th percentiles) for a given month of data as
\begin{equation}
g_{100}(k)= \max_{\substack{1\leq n \leq19, \, m\in\{1,2\} \\  k\tau\leq t < (k+1)\tau }} \,   g_m^n(t)
\label{eqn:max_series_derivation}
\end{equation}
where the $k \in \{0,1,...,\lfloor\frac{T}{\tau} \rfloor -1 \}$ are the temporal indices of the windows, $T$ is the amount of order book states in the given month of data at the maximum resolution of 10 seconds per state and $\lfloor \cdot \rfloor$ indicates the floor function. We define $g_{50}(k)$ as the time series of median gaps (that is, the 50th percentiles) in a similar way. This procedure is illustrated in figure \ref{fig:DiagSeriesWindowReduction}.

The returns are defined as 
\begin{equation}
    r(t) =  \ln \left( p(t) / p(t-1)  \right)
\end{equation}

where $p_t = (a_t^1 + b_t^1)/2$ is the middle price at time $t$. We chose to use middle prices in order to be consistent with previous works \cite{farmer2004,cristelli2010} that explore the relationship between gaps and returns.

From these we obtain the series of maximum returns which are defined as
\begin{equation}
    r_{100}(k) = \max_{k\tau \leq t < (k+1)\tau} r(t)
\end{equation}

with a similar definition for the median returns $r_{50}(k)$.

The reasons behind our choice to split the data into windows are two. The first is that by using the middle price to estimate the returns, a spurious and systematic dependence between the first gaps and the returns is introduced; by utilizing windows we can mitigate this effect to some extent. The second reason is that in a real financial market one would usually not be concerned with estimating the size of the return in the next instant of time, but rather the size of the maximum return within a given future time interval, which is precisely what causal discovery methods are equipped to do if a causal relationship is indeed present in the data.


\subsection{Gaps-Returns cross-correlation functions}

\begin{figure}[H]
\centering\includegraphics[width=0.5\textwidth]{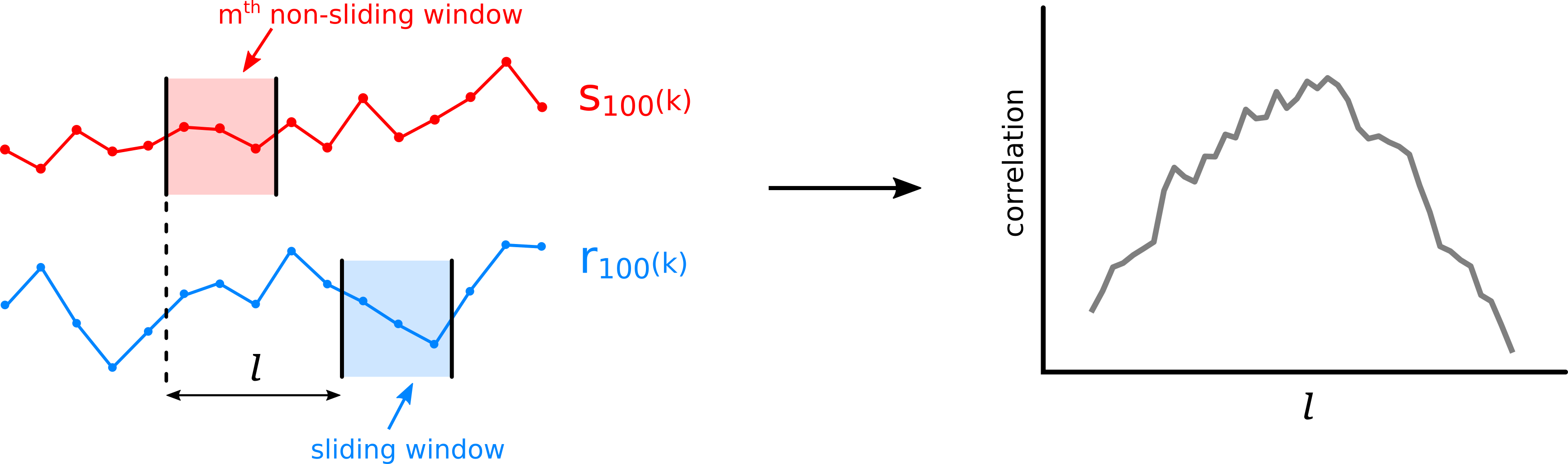}
\caption{Measurement of local correlation functions. A local correlation function between two series of maximums $s_{100}(k)$ and $r_{100}(k)$ obtained as shown in figure \ref{fig:DiagLocalCorrFunc} is computed by freezing a time window in $s_{100}(k)$ and calculating its correlations with a collection of lagged time windows in $r_{100}(k)$. By changing the location of the non-sliding window without intersecting its previous locations we end up with an ensemble of local correlation functions between $s_{100}(k)$ and $r_{100}(k)$, with one function for each non overlapping frozen window in $s_{100}(k)$.}
\label{fig:DiagLocalCorrFunc}
\end{figure}

Once the series of the maximum (median) gaps and returns are obtained, we compute the average of correlation functions obtained locally over small periods of time between the gaps and return statistics via the following procedure: the gap and return series $G(k)$ and $R(k)$ (here $G(k)$ stands for either $g_{100}(k)$ or $g_{50}(k)$, similarly for $R(k)$) are partitioned into $J$ disjoint windows of length $\widetilde{\tau}$. Next, a local correlation function is associated to the $j^{\text{\tiny th}}$ window in the returns, which is left fixed, by calculating its cross-correlations with a sequence of windows of length $\widetilde{\tau}$ located on $G(k)$. Each window on $G(k)$ has a lag in the index $k$ relative to the fixed $j^{\text{\tiny th}}$ window in $R(t)$, and is thus identified with its corresponding lag value $l\in\{-\widetilde{\tau},-\widetilde{\tau}+1,...,\widetilde{\tau}-1,\widetilde{\tau} \}$.  In this way, we end up with the local correlation function
\begin{equation}
C_{j}(l)=\frac{1}{\widetilde{\tau}\sigma_{G}\sigma_{R}}\sum_{i=t}^{t+\widetilde{\tau}}(R_{i}-\bar{R})(G_{i+l}-\bar{G})
\end{equation}
associated to the $j^{\text{\tiny th}}$ returns window, as shown in figure \ref{fig:DiagLocalCorrFunc}. Here $\bar{R}$ and $\bar{G}$ denote the averages and $\sigma_{G}$ and $\sigma_{R}$ the standard deviations of the returns and gaps within the $j^{\text{\tiny th}}$ window respectively.  Finally, once we obtain all the local correlation functions associated to each one of the $J$ windows in the disjoint partition of $R(k)$, we calculate the average local correlation function between $G(k)$ and $R(k)$ as 
\begin{equation}
    C(l) = \frac{1}{J} \sum_{j=1}^{N} C_{j}(l)
\end{equation}

By averaging over the ensemble of local correlation functions we make sure to highlight the geometrical properties that are present in a majority of the local functions and mitigate fluctuations around the average behavior.

\subsection{Granger causality tests}
It is said that a signal $X(t)$ Granger-causes another signal $Y(t)$ if a linear model including both the past of the explained variable $Y(t)$ and the causing variable $X(t)$ generates better predictions of the future of $Y(t)$ than a linear model using only the past of $Y(t)$. The performance of each model is evaluated by minimizing their sums of squared residuals (SSR) and comparing them, with a smaller quantity indicating better prediction capability of the model.
There are various ways in which a Granger causality test can be conducted, the particular implementation used in this work is as follows. A linear autoregressive model with a lag of $p$ (AR$(L)$) is fit to $Y(t)$:
\begin{equation}
    Y_{R}(t) = c_{R} + \sum_{l=1}^{L} \alpha_{l}Y(t-l) + \epsilon(t) 
\end{equation}

This linear regression of the explained variable is called the restricted model.

Another linear regression of $Y(t)$, called full model, is also fitted, but this time with the inclusion of past realizations of $X(t)$ as regressors:
\begin{equation}
    Y_{F}(t) = c_{F} + \sum_{l=1}^{L} \alpha_{l}Y(t-l) + \sum_{l=1}^{L} \beta_{l}X(t-l) + \eta(t) 
\end{equation}

With $\alpha,\beta,c_{R},c_{F}$ being the fitted coefficients of the models and $\epsilon (t), \eta (t)$ the error terms. The index $t$ runs from $t=1$ to $t=T$.

The hypothesis test formulated to assert whether $X(t)$ Granger causes $Y(t)$ is that the predictions of the full model are not statistically better than those of the restricted model. To determine this, a standard F-test is usually performed, with the statistic

\begin{equation}
    s = \frac{(SSR_{R}-SSR_{F})/L}{SSR_{F}/(T-2L-1)} 
\end{equation}

which follows, under a certain set of assumptions, an $F_{L,T-2L-1}$ distribution, thus allowing one to compute the probability of having observed a given difference in variances between both linear models.

Usually one finishes a Granger causality test with an affirmation (or negation) of the causal relationship between $X(t)$ and $Y(t)$, when the hypothesis was discarded (or not), under a significance level of $p$ (the $p$ value). In this work we rather based our conclusions directly on the statistic $s$ since this number serves as a proxy for the size of the effect that the inclusion of $X(t)$ has on the predictability of $Y(t)$, beyond merely answering in a binary way whether the presence of such a relationship is statistically defensible. The reason to proceed in this manner is that the distribution observed for $s$ differs from the expected F-distribution, which is unsurprising, since both the returns and gaps display fat tails. To make the effect sizes easier to interpret, we define them simply as
\begin{equation}
    s = \frac{(SSR_{R}-SSR_{F})}{SSR_{F}},
    \label{GrangerTestStatistic}
\end{equation}
where we do not normalize by the degrees of freedom of the $F$ distribution corresponding to the test statistic $s$, such modification does not alter the qualitative aspects of the results since we compare them with adequate controls. We will present the results of the Granger causality tests as a histogram of the $s$ values as they vary from one time window to the next and compare this with surrogate data which arise from a random shuffle of the original data, and should thus not show causal behavior.

We will use the nomenclature \textit{standard Granger} to refer to the tests that only take into consideration lagged versions of the suspect causal variable and \textit{instantaneous Granger} for the tests that include the causal variable with zero lag, as described in \cite{kirchgassner2007}. 

The instantaneous Granger causality tests are important because they will allow us to better distinguish whether the effects are essentially instantaneous or whether they can be used for forecasting. 

\subsection{Additive Noise Model causal discovery}

Causal discovery through Additive Noise Models (ANM)\cite{mooij2016} is a technique in which the explained variable  $Y$  is modeled as a function of the causing variable $X$ with an additive and independent noise term $\epsilon$ that acts as a proxy of the cumulative effect of every other latent causal variable that could affect the explained variable. So, the explained variable $Y$ is modeled as:
\begin{equation} 
Y = f(X) + \epsilon.  
\label{eqn:additive_noise}
\end{equation}
The central idea behind this technique is to assume that the data from two time series $\{(x_i,y_i),i=1,...,\tilde\tau\}$ are variates of a pair of random variables $X$ and $Y$, and that they are related via a one-way causal relationship $X\rightarrow Y$ in which $X$ directly affects $Y$ without the inverse relationship existing at the same time (that is, $Y$ affecting $X$) and assuming that any other causal effect that could come from causal variables other than $X$ can be represented as additive noise $\epsilon$, independent of $X$, as shown in equation \ref{eqn:additive_noise}.

To conduct a causality test by ANM one first needs to obtain an estimation of the function $f$ which we will denote $\hat f$ via a suitable regression method and then, from $\hat f$, obtain the residuals $\hat \epsilon_i$ as:
\begin{align} 
\hat \epsilon_i = y_i - \hat f(x_i).                          
\end{align}
After this is done, one tests the assumption that the variable $X$ is independent of $\epsilon$.  This can be done, given the regression analysis, by looking for a correlation between $x_i$ and the residuals $\hat\epsilon_i$.\\

In this work we made use of the Python package Causal Discovery Toolbox (CDT) \cite{kalainathan2019} which includes a module for ANM discovery. The CDT implementation uses Gaussian Process Regression \cite{roberts2013} to perform the regression analysis and a standard test for independence, namely, the Hilbert Schmidt Independence Criterion (HSIC)\cite{gretton2005} to measure the dependence between $X$ and $\epsilon$. Let us denote the magnitude of the HSIC statistic by $Z_{X\rightarrow Y}\geq 0$, this serves as an estimation of the possibly non-linear correlation between $x_i$ and the residuals $\hat\epsilon_i$. If $Z_{X\rightarrow Y}$ is big enough so that a correlation is detected, then it is discarded that $X$ and $Y$ follow an additive noise model in the direction $X\rightarrow Y$.

With these definitions at hand we now denote the effect size of the ANM causal discovery method as the score: 
\begin{align} 
S_{X,Y}=Z_{Y\rightarrow X}-Z_{X\rightarrow Y}                        
\end{align} which is also provided by the CDT implementation. The advantage of using the score $S_{X,Y}$ over $Z_{X\rightarrow Y}$  is that in order to validate an ANM in the direction $X\rightarrow Y$ it is not enough to obtain a low value of $Z_{X\rightarrow Y}$, but also a high value of $Z_{Y\rightarrow X}$ to be able to discard the causal relationship in the opposite direction. Indeed, merely having a low value of $Z_{X \to Y}$, does not distinguish between X causing Y and X being independent from Y. In this way the sign of $S_{X,Y}$ indicates the direction of the causal relationship if such a relationship is detected.

\section{Results and discussion}\label{section-results}
\subsection{Correlations over time}

\begin{figure}
   \includegraphics[width=0.7\linewidth]{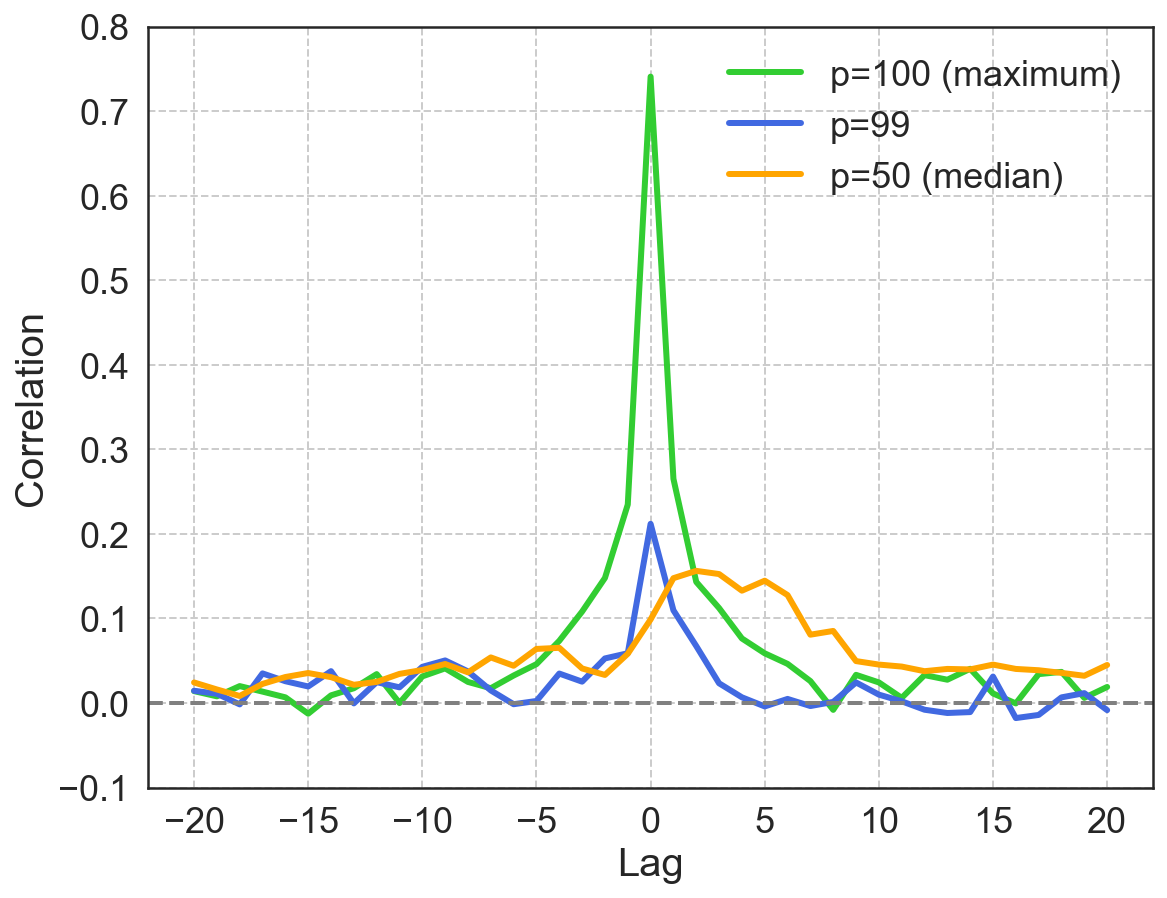}
   \caption{Average local correlations between the maximum ($p=100$), 99th percentile ($p=99$) and median ($p=50$) of the full gaps and returns. There is an abrupt change in the magnitude of the correlations as one considers percentiles other than the maximum. At a lg of zero, the correlations of 99th percentiles are almost 4 times smaller than those between the maximums.}
   \label{fig:CorFun_VolGap_Pearson_offset0}
\end{figure}

\begin{figure}
   \includegraphics[width=0.7\linewidth]{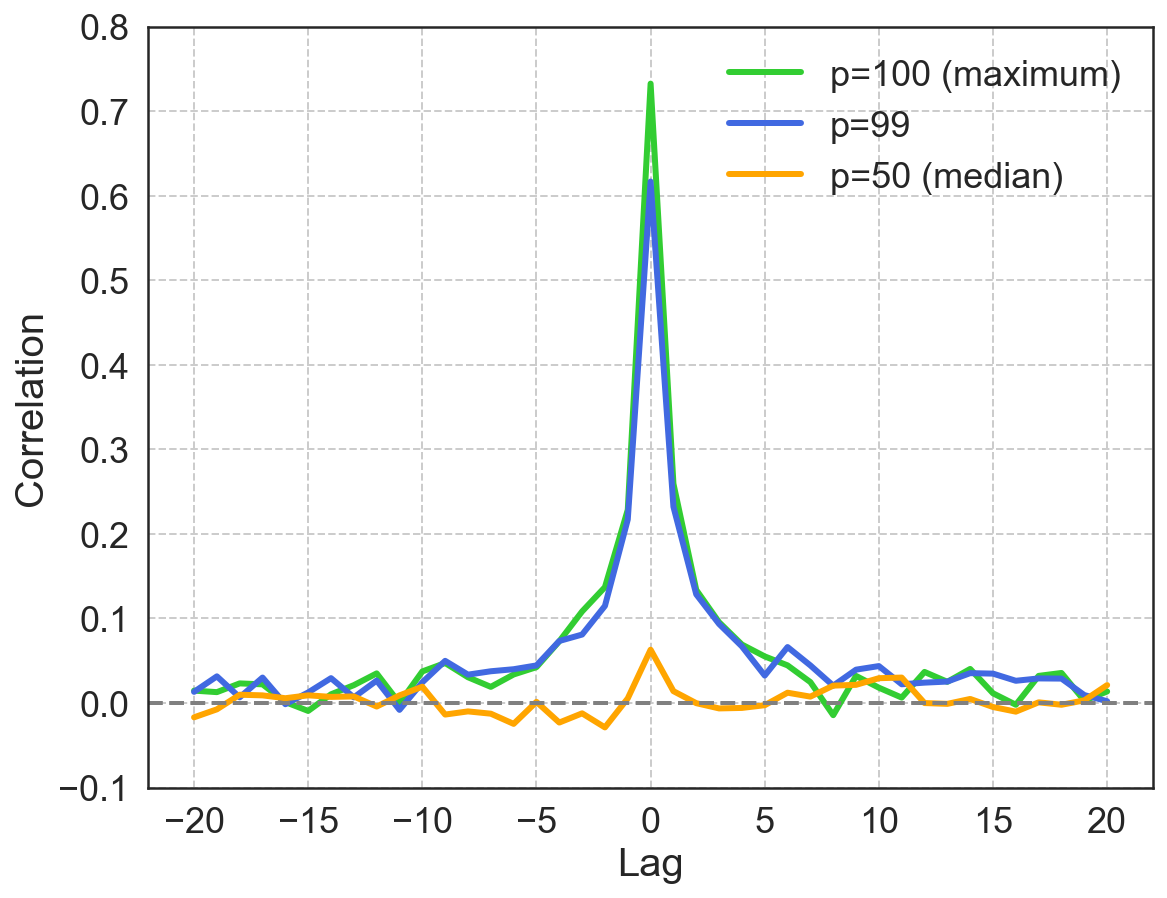}
   \caption{When only the first gaps ($g_1^1$ and $g_2^1$) are considered, the correlations between the maximums, 99th percentiles and median gaps and returns decrease gradually as lower percentiles are taken into consideration in sharp contrast with the correlations of the full gaps.}
   \label{fig:CorFun_Vol1stGap_Pearson} 
\end{figure}

We measured the correlations as a function of time between the maximum and median gaps and returns by splitting the order book multivariate series into disjoint windows of 60 time steps at a resolution of 10 seconds per time step. Next, we extracted from each of these windows the 50th and 100th percentiles (median and maximum), as described by equation \ref{eqn:max_series_derivation}. Once we derived the series of maximums and medians at the selected scale, we proceeded to calculate the average correlation function between them using a sliding window of 100 data points, where each data point is one of the maximums or medians extracted previously at a scale of 60 time steps. This procedure was  executed on each month of data separately. The average local correlation function between the percentile series was then computed. Figures \ref{fig:CorFun_VolGap_Pearson_offset0} and \ref{fig:CorFun_Vol1stGap_Pearson} show the correlation functions between the complete set of gaps and the returns and between only the first gaps and the returns, respectively. Note that the magnitude of the correlations is greatest when considering the maximum events, and quickly decreases in size as one considers lower percentiles, in particular, the correlation between the medians ($50_{th}$ percentile) is very small, compared to the correlation between the maximums.

For the computation of the correlation functions we considered two different scenarios: either we limited the set of gaps from which we extracted the maximum and medians to just the gaps in the first position, or we extracted them from the complete set of available gaps.
When only the first gaps are considered ($g_1^1$ and $g_2^1$), we observe that the maximum correlation between any pair of percentiles is attained at zero lag, having lower values for negative and positive lags. When the complete set of gaps is used, on the other hand, the $100_{th}$ and $99_{th}$ percentiles (the maximums) attain their maximum correlation at lag zero but the median (percentile 50) has its maximum at positive lags. Since the returns are the ones being fixed, the delayed position of the maximum correlation in the median implies that return statistics in the present are associated with future gap statistics. The reasons behind the delayed position of the maximum are unclear but they do not entail a causal relationship between gaps and returns.

\begin{figure}[H]
\centering\includegraphics[width=0.4\textwidth]{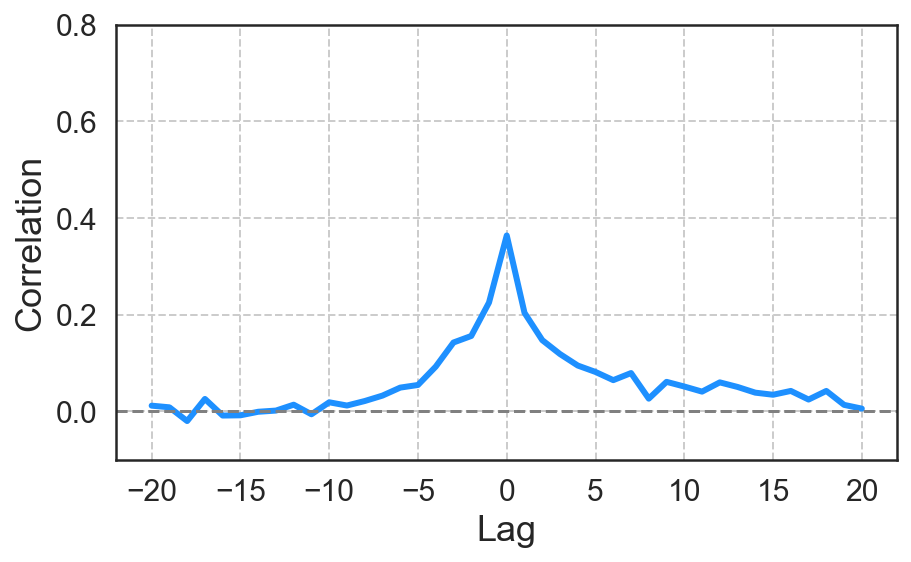}
\caption{Average local correlations between the median returns and maximum gaps. The correlation average is greater than in the previous case between the median of both variables. Return sizes ranging from the typical value (50th percentile) to the maximum value (100th percentile) are associated with the maximum gap size.}
\label{fig:MedRets_MaxGaps}
\end{figure}

The median size of gaps does not correlate as strongly to the median size of returns as do the maximums. By constrast, the correlation between the median return and the maximum gap is not as weak, as shown in figure  \ref{fig:MedRets_MaxGaps}, thus, both the typical and maximum size of returns are statistically associated with the maximum gaps.

These results also hold for window sizes different from 100 time steps.
\subsection{Granger causality tests}

The fact that the correlation functions do not vanish immediately for lags different from zero is compatible with a causal relationship but insufficient to demonstrate it.
A direct causality measure such as Granger causality can be used in order to distinguish whether the relationships observed are consistent with the idea of causal effects or are better explained as correlations. 

\begin{figure}[h]
\centering
\includegraphics[width=0.35\textwidth]{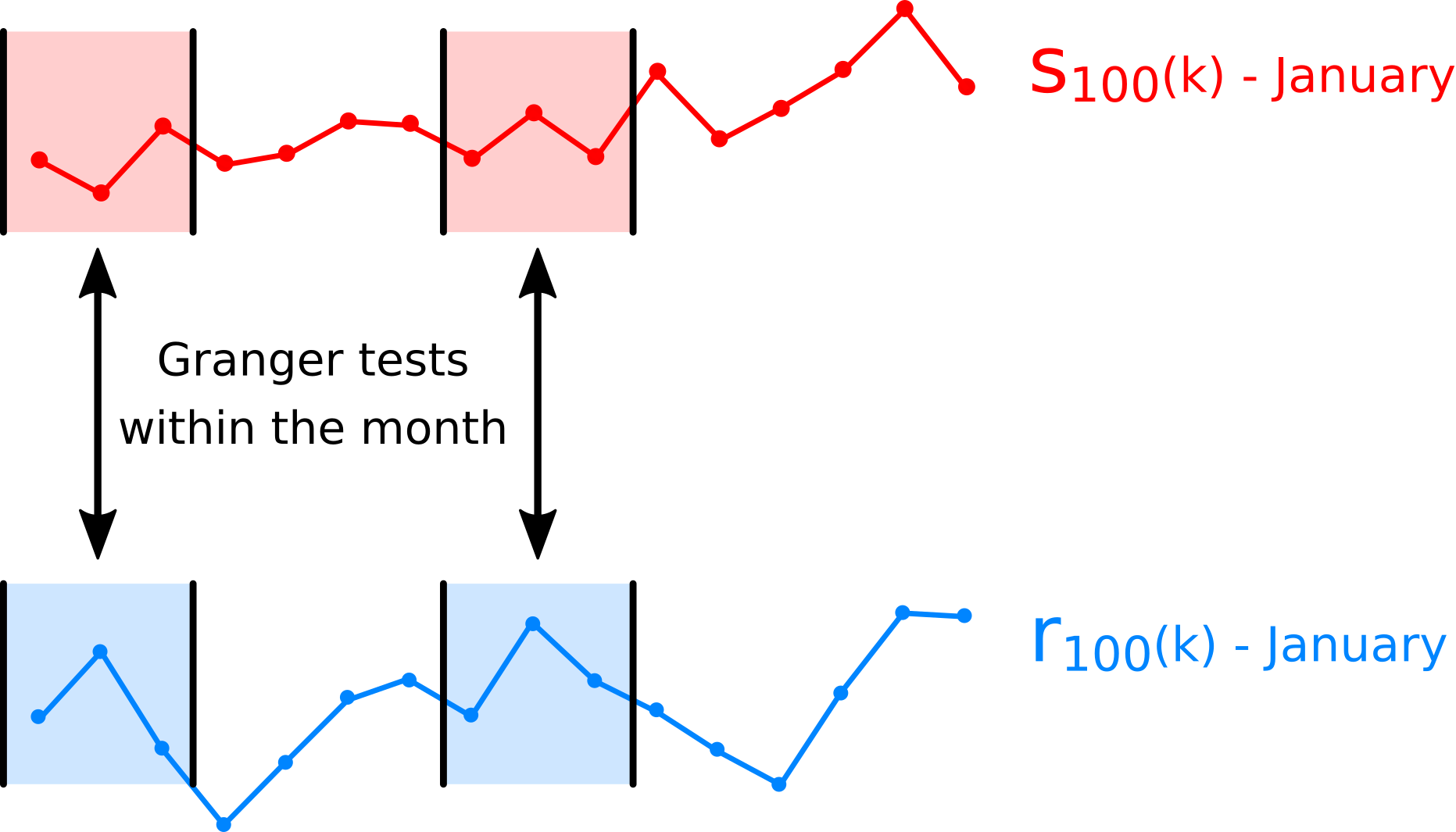}
\caption{Illustration of the partitioning of each month of data into windows to increase the sample size of the Granger tests. A given month (January in the figure) of a series obtained as described in \ref{fig:DiagSeriesWindowReduction} is partitioned into disjoint windows of $\tilde \tau = 500$ time steps and a Granger test is conducted between each pair of windows. This procedure is performed on all 14 months of data available and the results are collected to compute the various distributions of Granger causality scores of figures \ref{fig:GC_standard_window501_fullgaps} and \ref{fig:GC_instantfull_window501_fullgaps}.}
\label{fig:WindowsGranger}
\end{figure}

For this purpose, aside from the correlation functions between the gap and return percentiles, we conducted simple Granger causality tests in both causal directions: gaps to returns and vice-versa. 
The Granger tests were conducted at four different timescales ($\tau = 30,60,90$ or $120$ steps at 10 seconds per step) by dividing each month of the series $r_{50},r_{100},g_{50}$ and $g_{100}$ into disjoint windows with a length of $\widetilde{\tau} = 500$ time steps, as described in figure \ref{fig:WindowsGranger}. This segmentation has the purpose of increasing the sample size of causality scores, which would be too small if we only conducted the tests on the 14 months of available data.  It is important to remark that within the series $r_{50},r_{100},g_{50}$ and $g_{100}$, each time step corresponds to either a maximum or a median gap (return) taken over a scale $\tau$ as mentioned before.

The maximum lag over the windows (indexed by $k$) was chosen via the Bayesian information criterion\cite{gordon2015}(BIC), which was performed on the auto-regressive models of each series with lag values ranging from $l=1$ to $l=10$. We chose the value of $l$ that achieved the minimum median BIC across this range of lags and across all values of $\tau$.

\begin{figure}[h]
\centering
\includegraphics[width=0.45\textwidth]{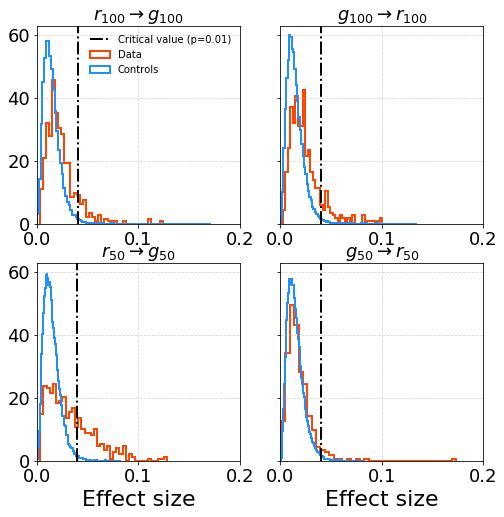}
\caption{Effect sizes for the Granger causality tests. Red corresponds to the tests conducted on the original data and blue to the tests on shuffled surrogates. The dotted black line marks the critical value corresponding to a p-value of p=0.01. Since the random shuffling destroys any time dependence between time series, the shuffled surrogates represent the null hypothesis of no causality between gaps and returns. The distributions of effect sizes of the tests conducted on the real data are very similar to those conducted on the surrogates both in the shape and support of their distributions, in accordance to a lack of a Granger causal relationship between these two signals.}
\label{fig:GC_standard_window501_fullgaps}
\end{figure}

The effect sizes, corresponding to the non-normalized $s$ statistic mentioned previously (see eq \ref{GrangerTestStatistic}), are shown in the figure \ref{fig:GC_standard_window501_fullgaps}, where the red histograms correspond to the results of the tests on the original data while the blue ones to the tests of randomly shuffled copies of the data, used as controls.
Except for relationship between the median returns and gaps, the distributions of causality scores of the controls and the real data have very similar shapes and supports and very few tests achieved statistical significance under a critical value of p=0.01, with only 12\% of tests surpassing the critical value in the case of the relationship $g_{100}\rightarrow r_{100}$ and 10\% in the case of $r_{100}\rightarrow g_{100}$.

In the standard Granger tests as well as in the ones that include instantaneous effects, no asymmetry between the effect sizes of both causal directions $g_{100}\rightarrow r_{100}$ and $r_{100}\rightarrow g_{100}$ is evident, as such, it is roughly equally likely to measure a given effect size in either direction.

\begin{figure}[h]
\centering
\includegraphics[width=0.45\textwidth]{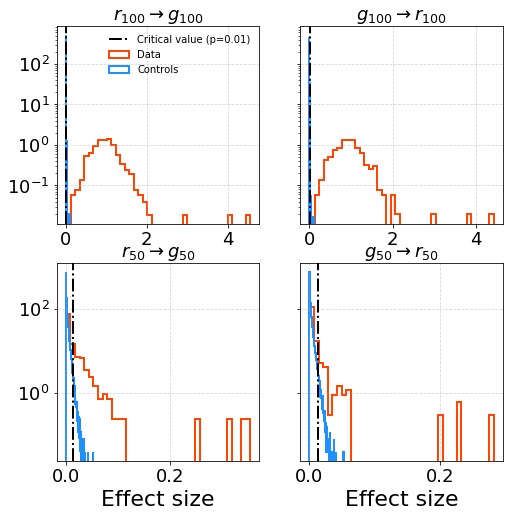}
\caption{Effect sizes for the Granger causality tests with instantaneous regression term. This time the distributions of effect sizes between the original data and the controls are very different when considering the maximums in both causal directions. The median pairs still are remarkably similar.}
\label{fig:GC_instantfull_window501_fullgaps}
\end{figure}

When instantaneous windows of the causal variable are added in the linear regressions of the explained variables, that is, when the minimum lag in the regressive models is zero, a very different result is observed, as shown in figure \ref{fig:GC_instantfull_window501_fullgaps}. Although the controls and the real data distributions are similar for the medians, for the maximums the shapes of the distributions become very dissimilar and the effect sizes blow up, with 100\% of the tests achieving statistical significance in both causal directions. This suggests that we do not detect a true causality allowing to forecast return sizes reliably from the knowledge of gaps.

\begin{table}
\centering
\begin{tabular}{cccc} \toprule
                   &      Standard   &   Instant \\ 
{$g_{100}\rightarrow r_{100}$}  & 0.12    &   1.00    \\
{$r_{100}\rightarrow g_{100}$}  & 0.10    &   1.00    \\
{$g_{50}\rightarrow r_{50}$}    & 0.04    &   0.11    \\
{$r_{50}\rightarrow g_{50}$}    & 0.30    &   0.27    \\ \\ 
\end{tabular}

\caption{Proportions of statistically significant causality scores in the Granger tests for the null hypothesis of no causal effect under a p-value of p=0.01. The standard tests throw a very small number of significant tests. When we consider the inclusion of instantaneous windows in the linear models of the test, the causality scores blow up, with every single test surpassing the critical region, in the case of the maximum gaps/returns. The medians remain with low percentages of significant tests.}
\label{tab:Significancias_Estandar_Granger}
\end{table}

\subsection{Additive Noise Models tests}
Now we describe the results of the application of causal discovery by means of Additive Noise Models. Since ANM tests do not inherently consider the chronological order between the causal and explained variables as Granger causality does, we performed tests between the gap and return percentile series considering no lag between them, which corresponds to instantaneous causal relationships, and between lagged versions of the series, mimicking the chronological order taken of the standard Granger causality tests. 
There are two key differences between this way of forcing chronological order into the ANM tests and the way in which Granger causality naturally considers time via the lags in its regressive models. The first difference is that for a given maximum lag value $\tau\in \mathbb{N}$, the test of Granger takes into account lagged observations ranging from 1 to $\tau$ time-steps in the past while the ANM test will only consider a lag of $\tau$ time-steps. The second difference is that in the case of non-zero lags, only the positive results have meaning since negative causality scores would correspond to causal effects from the series located in the future to the series located in the past.

\begin{figure}[h]
\centering
\includegraphics[width=0.4\textwidth]{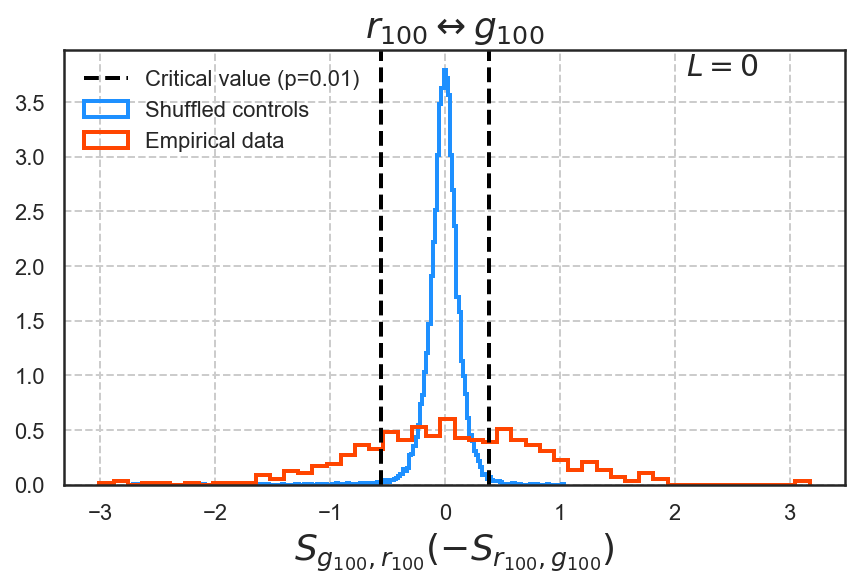}
\caption{Effect sizes for the ANM tests applied to the maximum returns and gaps with a lag value of $L=0$. A positive effect size $(S_{g_{100},r_{100}})$ corresponds to a causal effect from the gaps size to the returns size and a negative one $(-S_{r_{100},g_{100}})$ to an effect from the returns to the gaps. The dotted black lines mark the two-tailed critical values corresponding to a p-value of p=0.01. The shapes of the control and data distributions are very dissimilar and a significant proportion of the cases fall far away from the tails of the controls.}
\label{fig:ANM_Maxes_L0_RG-GR}
\end{figure}

\begin{figure}[h]
\centering
\includegraphics[width=0.4\textwidth]{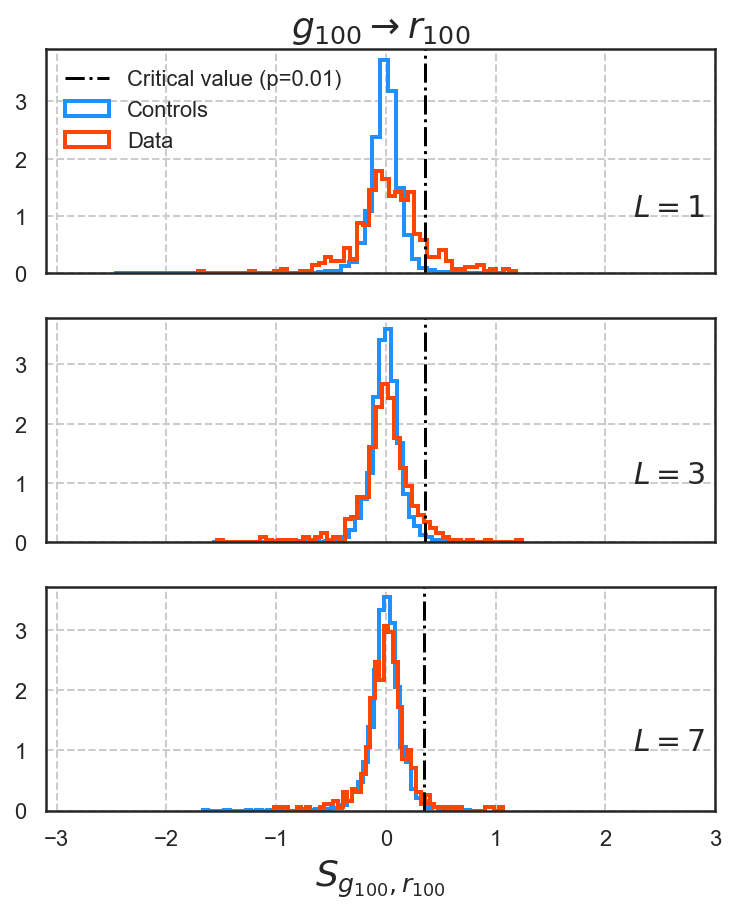}
\caption{Effect sizes for the ANM tests for the causal relationship $g_{100}\rightarrow r_{100}$ with positive lag values. Only the positive causality scores correspond to a detection of causal effects since negative values would imply that the future return sizes have an effect on past gap sizes. The dotted black lines mark the one-tailed critical values corresponding to a p-value of p=0.01. As opposed to the case with zero lag, now the shapes of the control and data distributions are similar and only a small fraction of the tests fall far from the tails of the controls.}
\label{fig:ANM_Maxes_L1_L3_L7_GR}
\end{figure}

\begin{figure}[h]
\centering
\includegraphics[width=0.4\textwidth]{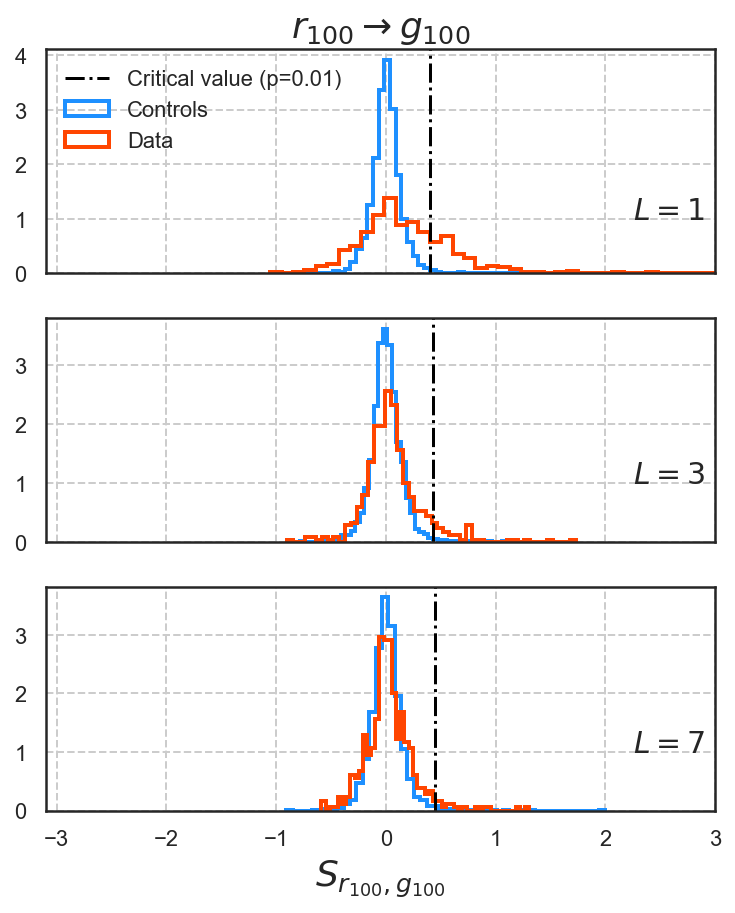}
\caption{Effect sizes for the ANM tests for the causal relationship $r_{100}\rightarrow g_{100}$ with positive lag values. Only the positive scores are associated to realistic causal effects. The dotted black lines mark the one-tailed critical values corresponding to a p-value of p=0.01. Again, the shapes of the control and data distributions are similar and few tests fall outside the critical zone under $p=0.01$.}
\label{fig:ANM_Maxes_L1_L3_L7_RG}
\end{figure}

The results of the ANM tests, shown in figures \ref{fig:ANM_Maxes_L0_RG-GR},\ref{fig:ANM_Maxes_L1_L3_L7_GR}, \ref{fig:ANM_Maxes_L1_L3_L7_RG} and table \ref{tab:Significancias_Estandar_ANM}, are in agreement with the Granger causality tests in both cases: the case where no lag is considered, which shows the highest deviation of causality scores with respect to the control scores distribution, and the lagged cases, where the distribution of scores becomes similar to the control distribution, even when considering a lag of a single time step. 

For the causal relationships between the medians, no statistically significant effect is found independently of the value of the lag since the ANM causality scores are totally contained within the values of the control distribution. A peculiarity of the causality scores between the medians is that these are all negative for the data and the controls. A possible explanation of this phenomenon is given in the discussion section.

We repeated the Granger and ANM tests between the gaps and two different definitions of volatility: as the average of absolute logarithmic returns and as the standard deviation of logarithmic returns. We obtain the same qualitative results in those tests and could not discard the null hypothesis of absence of a causal relationship.

\begin{table}
\centering
\begin{tabular}{ccccc} \toprule
                        &       $L=0$     &   $L=1$   &  $L=3$ & $L=7$ \\ 
{$g_{100}\rightarrow r_{100}$}  & 0.35    &   0.12    &   0.05 & 0.03 \\
{$r_{100}\rightarrow g_{100}$}  & 0.21    &   0.26    &   0.08 & 0.03\\\\ 
\end{tabular}
\caption{Proportions of statistically significant causality scores in the ANM tests at a level of $p=0.01$. We see that mostly when the gaps and returns series are considered without a lag a considerable proportion of tests are statistically significant.}
\label{tab:Significancias_Estandar_ANM}
\end{table}

\section{Discussion}

\subsection{Simultaneous causal relationships}

There are statistical associations between the maximum gaps and maximum returns and the maximum gaps and median returns that do not immediately die off as evidenced by the correlation functions shown in figures \ref{fig:CorFun_Vol1stGap_Pearson} and \ref{fig:MedRets_MaxGaps}. With the exception of the median gaps and returns, these correlations attain their highest magnitude at zero lag in a similar way to that in which the causality tests attain the highest scores when the observations are taken into account with the inclusion of instantaneous windows.

In such scenario, all the tests for Granger causality become statistically significant in both causal directions: $g_{100}\rightarrow r_{100}$ and $r_{100}\rightarrow g_{100}$, which only indicate a strong simultaneous relationship between maximum gaps and maximum returns but without pointing out a particular direction (figure \ref{fig:GC_instantfull_window501_fullgaps}), this indicates that there is no direction to the ``causality'' observed. Furthermore, the ANM tests differ significantly at both tails of the causality scores distribution with respect to the control distribution (figure \ref{fig:ANM_Maxes_L0_RG-GR}).

So, with the inclusion of instantaneous time periods, the Granger tests fully reject the null hypothesis of no causal effect. Similarly, although the ANM tests do not fall completely outside of the control distribution support, they make it difficult to accept the null hypothesis since more than half of the tests reach statistical significance with a difference of 1.66 times more tests favoring the direction $g_{100}\rightarrow r_{100}$ than the opposite direction $r_{100}\rightarrow g_{100}$.

These results imply that there is a link between the size of the maximum gap and maximum return of middle price within the same time interval $[t,t+\tau]$. Therefore, the correlation functions and the causality tests with instantaneous windows are consistent with the results shown in previous literature\cite{farmer2004,cristelli2010} since they show that the existence of large gaps in the book is associated with the generation of large returns, when returns are defined as mid-price returns.

\begin{figure}[h]
\centering
\includegraphics[width=0.4\textwidth]{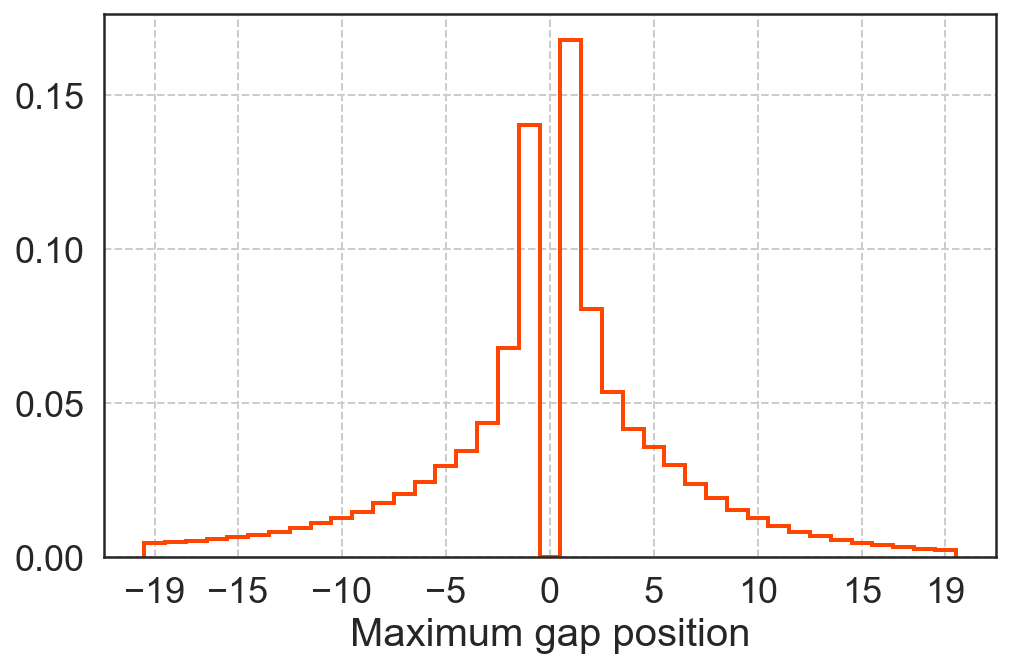}
\caption{Distribution of the positions at which the maximum gaps is located. The sign of the positions correspond to the sides od the book: positive for the asks and negative for the bids. That is, a value of $-1$ indicates the first gap in the bids and correspondingly, a value of $1$ indicates the first gap in the asks. It can be seen that is unlikely to find the maximum gap beyond the fifth position, with a probability of roughly 30\% of being located in the first position of either side.}
\label{fig:MaxGapDistro}
\end{figure}

There are, however, caveats with the positive results of the instantaneous causality tests: since the returns were defined as a function of the middle price, an artificial dependence of the returns on the first gaps of the order book is induced. The maximum gap is usually the first one: in the data analyzed around 30\% of the maximum gaps are located at the first level, as shown in figure \ref{fig:MaxGapDistro}. Appearance or disappearance of solely a first gap due to any reason (limit orders placements, cancellations or transactions) implies a change of mid-price proportional to the size of that first gap. Thus, maximum returns within the time interval $[t,t+\tau]$ often correspond to changes when the first gap is also the maximum gap. This correspondence naturally rises a strong statistical relation between the sizes of the maximum returns and the first gaps. 

For these reasons we must be very careful when interpreting the positive results of the instantaneous causality tests. 

\subsection{Time lagged causal relationships}

With respect to the standard causal scenario where lagged versions of the observations are considered, both the Granger and ANM tests throw mostly non-significant results (\ref{fig:GC_standard_window501_fullgaps}, \ref{fig:ANM_Maxes_L1_L3_L7_GR}, \ref{fig:ANM_Maxes_L1_L3_L7_RG} and table \ref{tab:Significancias_Estandar_ANM}). Among all tests, only few of them achieve statistical significance, so their distributions for each case are very similar to their respective control distributions. This similarity is important since randomly permuted surrogates keep the same moments and distribution as the original data but any structure that depends on the temporal order is destroyed\cite{lancaster2018}, which turns them into a very aggressive control scenario. So, even when we destroy all chronological structure we get very similar causality score distributions to the ones obtained in the original data, a result that strongly suggests the absence of a cause and effect relationship.

\begin{figure}[h]
\centering
\includegraphics[width=0.5\textwidth]{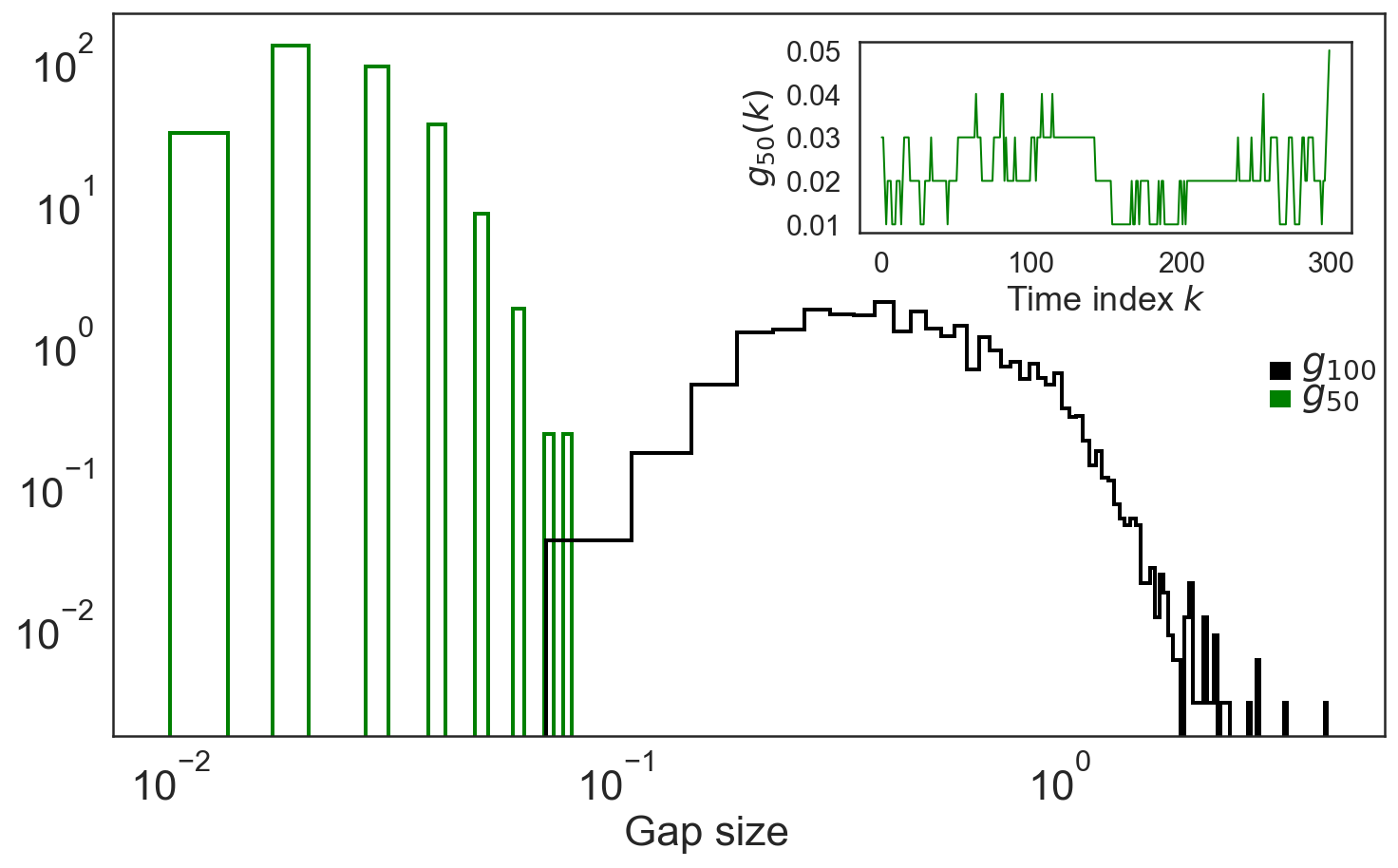}
\caption{Distribution of gap sizes for the median and maximum gaps. It can be seen that the maximum gaps explore a bigger support than the median gaps and that the cuantization of the gaps induced by the minimum price change of $0.01$ USD is very present in the median gaps. The inset shows a fragment of the median gaps as a function of time, and, as expected from the support of distribution, the time series is heavily discretized.}
\label{fig:Median_Gaps_Distribution}
\end{figure}

As mentioned earlier, all the ANM tests for the medians resulted non statistically significant since the support of the distribution of causality scores from the data is completely contained within the support of the controls distribution, with the peculiarity that in all tests, both in the data and the controls, the test statistic is negative. This would suggest the causal direction $r_{50} \rightarrow g_{50}$ if the tests were statistically significant, which is not the case. Aside from that, the correlation function between the median gaps and returns shown in Figure \ref{fig:CorFun_VolGap_Pearson_offset0} and the distribution of Granger causality test effect sizes in the direction $r_{50}\rightarrow g_{50}$ shown in Figure \ref{fig:GC_standard_window501_fullgaps} could suggest the existence of a small causal effect on the median size of the gaps stemming from the median returns. We think this is not the case since the median gaps usually reside above the first levels in the order book, where the market activity rarely consumes volume. Although we do not have definite answer to this issue, a possible explanation could be that this is an artifact of the quantized nature of the median gap sizes along with the fact that the median gaps explore only a small region of possible values, as opposed to the maximum gaps that explore a wide range of values from a few cents to several dollars, as seen in Figure \ref{fig:Median_Gaps_Distribution}.


Taken together, the nullity of the standard causality tests suggest that the size of the largest return found within the time interval $[t+\tau+1,t+2\tau + 1]$ can not be causally linked to the size of the largest gap found in the preceding interval $[t,t+\tau]$, again, to the extent of the causality definitions assumed by the techniques we used and the time scales we explored. 

It does not appear to be the case that the results found previously in the literature explaining the differences in the tails of the returns distributions of different companies as a consequence of the granularity of their respective order books apply when the granularity of a single order book is tracked over time. This could be a particularity the bitcoin/dollar market so a replication of this study on the stock market would be of use to test for this possibility.

\section{Conclusions}
The work presented here broadens previous works on the relationship between the granularity of the order book and the returns across a population of order books in which we took into consideration a single instance of an order book and tracked the evolution of the gap sizes across time with the aid of causal discovery methods in an attempt to test the hypothesis of the presence of large gaps in the book as a primary source of large returns.
We have shown for the Bitcoin/USD trading platform BTC-e, that the evolution across time of large price fluctuations does not seem to be a consequence of fluctuations in the size of the gaps present in the book in the sense that having access to the present state of the gaps appears to be of little use to create estimates of the future size of maximum returns in future windows of time. On the other hand, when the same time interval is considered in both the series of gaps and returns, we found that the the size of the maximum gap is an excellent predictor of the size of the maximum return. These results could be a particularity of the financial market that we studied or of the minimum time scales imposed by the data but at the very least they impose restrictions on the temporal horizon and precise definition of causal relationship that one can expect between the granularity of the book and the tail properties of the returns. Knowing such restrictions can be of help to better guide the sources of information considered in the creation of volatility forecasting models. A replication of our tests on different markets, at smaller timescales and possibly with complementary causal discovery tools would be of much use to obtain stronger constraints on the conclusion that knowledge of the state of the gaps in the order book does not seem to benefit the estimation of future price fluctuations.

\section*{Acknowledgments}

We would like to thank Hernán Larralde for his helpful discussion and feedback. The authors gratefully acknowledge the financial support from CONACyT, Grant Nos. Proyecto Fronteras 201 and CONACyT CB 254515, and the UNAM PAPIIT research, Grant No. IN113620. Roberto Mota Navarro thanks CONACyT for a Ph.D. scholarship.
 
\newpage	
\nocite{*}
\bibliographystyle{unsrt}

\appendix

\bibliography{references} 

\begin{thebibliography}{10}

\bibitem{biais1995}
Bruno Biais, Pierre Hillion, and Chester Spatt.
\newblock An empirical analysis of the limit order book and the order flow in
  the paris bourse.
\newblock {\em the Journal of Finance}, 50(5):1655--1689, 1995.

\bibitem{farmer2005}
J~Doyne Farmer, Paolo Patelli, and Ilija~I Zovko.
\newblock The predictive power of zero intelligence in financial markets.
\newblock {\em Proceedings of the National Academy of Sciences},
  102(6):2254--2259, 2005.

\bibitem{goldstein2000}
Michael~A Goldstein and Kenneth~A Kavajecz.
\newblock Eighths, sixteenths, and market depth: changes in tick size and
  liquidity provision on the nyse.
\newblock {\em Journal of Financial Economics}, 56(1):125--149, 2000.

\bibitem{gould2013}
Martin~D Gould, Mason~A Porter, Stacy Williams, Mark McDonald, Daniel~J Fenn,
  and Sam~D Howison.
\newblock Limit order books.
\newblock {\em Quantitative Finance}, 13(11):1709--1742, 2013.

\bibitem{schmitt2012}
Thilo~A Schmitt, Rudi Sch{\"a}fer, Michael~C M{\"u}nnix, and Thomas Guhr.
\newblock Microscopic understanding of heavy-tailed return distributions in an
  agent-based model.
\newblock {\em EPL (Europhysics Letters)}, 100(3):38005, 2012.

\bibitem{cristelli2010}
Matthieu Cristelli, V~Alfi, L~Pietronero, and A~Zaccaria.
\newblock Liquidity crisis, granularity of the order book and price
  fluctuations.
\newblock {\em The European Physical Journal B}, 73(1):41--49, 2010.

\bibitem{farmer2004}
J~Doyne~Farmer, Laszlo Gillemot, Fabrizio Lillo, Szabolcs Mike, and Anindya
  Sen.
\newblock What really causes large price changes?
\newblock {\em Quantitative finance}, 4(4):383--397, 2004.

\bibitem{gabaix2003}
Xavier Gabaix, Parameswaran Gopikrishnan, Vasiliki Plerou, and H~Eugene
  Stanley.
\newblock A theory of power-law distributions in financial market fluctuations.
\newblock {\em Nature}, 423(6937):267--270, 2003.

\bibitem{kirchgassner2007}
Gebhard Kirchg{\"a}ssner and J{\"u}rgen Wolters.
\newblock {\em Introduction to modern time series analysis}.
\newblock Springer Science \& Business Media, 2007.

\bibitem{mooij2016}
Joris~M Mooij, Jonas Peters, Dominik Janzing, Jakob Zscheischler, and Bernhard
  Sch{\"o}lkopf.
\newblock Distinguishing cause from effect using observational data: methods
  and benchmarks.
\newblock {\em The Journal of Machine Learning Research}, 17(1):1103--1204,
  2016.

\bibitem{kalainathan2019}
Diviyan Kalainathan and Olivier Goudet.
\newblock Causal discovery toolbox: Uncover causal relationships in python.
\newblock {\em arXiv preprint arXiv:1903.02278}, 2019.

\bibitem{roberts2013}
Stephen Roberts, Michael Osborne, Mark Ebden, Steven Reece, Neale Gibson, and
  Suzanne Aigrain.
\newblock Gaussian processes for time-series modelling.
\newblock {\em Philosophical Transactions of the Royal Society A: Mathematical,
  Physical and Engineering Sciences}, 371(1984):20110550, 2013.

\bibitem{gretton2005}
Arthur Gretton, Olivier Bousquet, Alex Smola, and Bernhard Sch{\"o}lkopf.
\newblock Measuring statistical dependence with hilbert-schmidt norms.
\newblock In {\em International conference on algorithmic learning theory},
  pages 63--77. Springer, 2005.

\bibitem{gordon2015}
Rachel~A Gordon.
\newblock {\em Regression analysis for the social sciences}.
\newblock Routledge, 2015.

\bibitem{lancaster2018}
Gemma Lancaster, Dmytro Iatsenko, Aleksandra Pidde, Valentina Ticcinelli, and
  Aneta Stefanovska.
\newblock Surrogate data for hypothesis testing of physical systems.
\newblock {\em Physics Reports}, 748:1--60, 2018.

\bibitem{diviyan2020}
Diviyan Kalainathan, Olivier Goudet, and Ritik Dutta.
\newblock Causal discovery toolbox: Uncovering causal relationships in python.
\newblock {\em Journal of Machine Learning Research}, 21(37):1--5, 2020.

\end{thebibliography}

\end{document}